\documentstyle[aps,twocolumn,psfig]{revtex}
\begin{document}
\draft

\title{Quantum Measurement and Entropy Production}
\author{Paolo Grigolini$^{1,2,3}$, Marco G. Pala$^{1}$, Luigi
Palatella$^{1}$}
\address{$^{1}$Dipartimento di Fisica dell'Universit\`{a} di Pisa and
INFM,
Piazza Torricelli 2, 56127 Pisa, Italy}
\address{$^{2}$Istituto di Biofisica CNR, Area della Ricerca di Pisa,
Via Alfieri 1, San Cataldo 56010 Ghezzano-Pisa,  Italy}
\address{$^{3}$Center for Nonlinear Science, University of North Texas,\\
P.O. Box 5368, Denton, Texas 76}
\date{\today}
\maketitle

\begin{abstract}
We study the time evolution of a quantum system 
without classical counterpart,
undergoing a process of entropy increase due to the environment
influence. We show that if the environment-induced decoherence is
interpreted in terms of wave-function collapses, a symbolic sequence
can be generated. We prove that the Kolmogorov-Sinai entropy of this
sequence coincides with rate of von Neumann entropy increase.

\end{abstract}

\pacs{03.65.Sq,05.20.-y,03.65.Bz}

According to Landau and Lifshitz\cite{landau1,landau2} the foundation
of the second law might lie in the processes of quantum measurement.
This point of view, still under the form of a plausible conjecture,
has been recently reformulated by Srivastava, Vitiello and
Widom\cite{widom}.
The authors of this interesting paper note that the von Neumann
entropy, which is kept constant by the unitary transformation of
quantum mechanics, increases as a consequence of the von Neumann projective
measurement, and consequently as an effect of the occurrence of a
quantum measurement. Their approach makes manifest the concept of
heat and work during the measurement process. In summary,
they prove that the von Neumann entropy
expressed in terms of the von Neumann projected density matrix, as a
consequence of  quantum measurement, becomes indistinguishable from
the physical entropy.

The present letter focuses on a different aspect of the same fundamental
issue. This has to do with the relation between physical entropy and
Kolmogorov-Sinai (KS) entropy\cite{kolmogorov,sinai}. This latter form
of entropy is actually a property of a classical
trajectory\cite{beck,dorfman}. The
classical phase space is divided into cells, the cells are labelled
with symbols, and a trajectory running in this phase space creates a
symbolic sequence. Finally, using the KS entropy prescription this
trajectory is assigned the value $h_{KS}$ that can be interpreted as
a rate of entropy increase.
In the recent past many papers have been devoted to the discussion of
the possible connection between the KS entropy and the physical
entropies~\cite{zurek,pattanayak1,pattanayak2,pattanayak3,pattanayak4,miller1,miller2,bhattacharya,vulpiani,latora}.
Of special relevance for the discussion of this paper is the work of
Latora and Baranger\cite{latora}. These authors study the time
evolution of the physical entropy moving from an out of equilibrium
initial condition and prove that three distinct time regimes exist:
An initial regime of transition, an intermediate regime of linear
increase and a saturation regime. The rate of entropy increase in the
intermediate regime is proved by them to coincide with the KS entropy.
Results of the same kind have been derived by Pattanyak\cite{pattanayak4}
along lines that essentially adopt a perspective originally advocated
by Zurek and Paz\cite{zurek}.
In a sense, the perspective of Zurek and Paz is the same as the
perspective of Refs.\cite{landau1,landau2,widom} if we interpret the
influence of the environment as a nature-made form of
measurement\cite{carmichael,gisin,dalibard}.
However, the main limitation of all these papers is given by the fact that
the adopted approach works only when the system studied has a classical
counterpart. Thus, the corrrespondence between the original 
conjecture  of Landau and
Lifsthiz\cite{landau1,landau2} and physical entropy\cite{widom}, on
one hand, and physical entropy
and KS entropy
~\cite{zurek,pattanayak1,pattanayak2,pattanayak3,pattanayak4,miller1,miller2,bhattacharya,vulpiani,latora},
on the other
hand,
is only partially established.

The main purpose of this letter is to fill this gap and to show that
the KS entropy shows up even throughout the environment-done
measurement process of systems which do not have a classical analog. 
Let us consider the quantum mechanical Hamiltonian:
\begin{equation}
    H = (|1\rangle\langle2| + |2\rangle\langle1|)V.
    \label{hamiltonian}
    \end{equation}
    We imagine this system as consisting of two distinct sites, for
    instance two wells of a lattice coupled to one another by a
    tunneling process with rate $\hbar/(2V)$. The solution
    corresponding to the initial condition $|\psi(0)\rangle = |1\rangle$
    reads
    \begin{equation}
        |\psi(t)\rangle = \cos(\frac{V t}{\hbar}) |1\rangle - 
        i \sin(\frac{V t}{\hbar}) |2\rangle.
        \label{harmonic}
        \end{equation}
      In the presence of the interaction with an environment causing
      decoherence, with the rate $2 \sigma/\hbar$, it is convenient to use
      the statistical density matrix. This physical condition is thus
      expressed by
     
      \begin{eqnarray}
\label{tutte}
      \dot{\rho}_{11}(t)& = & -i\frac{V}{\hbar}(\rho_{21}-\rho_{12}) 
\nonumber \\
      \dot{\rho}_{12}(t)& = & -i\frac{V}{\hbar}(\rho_{22}-\rho_{11}) -
      \frac{2\sigma}{\hbar} \rho_{12}(t) \nonumber  \nonumber \\
      \dot{\rho}_{22}(t)& = & -i\frac{V}{\hbar}(\rho_{12}-\rho_{21}) 
\nonumber \\
      \dot{\rho}_{21}(t)& = & -i\frac{V}{\hbar}(\rho_{11}-\rho_{22}) -
      \frac{2\sigma}{\hbar} \rho_{21}(t)   .
      \end{eqnarray}
      In this letter we limit our attention to the
      case:
      \begin{equation}
	   V << \sigma
	   \label{frequentmeasurement}.
	   \end{equation}
Note that the condition of Eq.(\ref{frequentmeasurement}) makes
it possible for us to define the time region where the intermediate
regime of KS type is expected to be located. The time
$ \tau \equiv \hbar/\sigma$ is evidently the time at which 
entropy starts increasing. 
The time evolution of $\delta(t) \equiv
\rho_{11}(t) - \rho_{22}(t)$ is easily proven\cite{grigo} to be
equivalent to that of a damped harmonic oscillator with
frequency $2 V/\hbar$ and damping $2 \sigma/ \hbar$. The condition of
Eq.(\ref{frequentmeasurement}) sets the overdamping condition, and
consequently establishes the time scale $T_{relax} \equiv 
(\hbar \sigma)/2 V^{2}$ as
the time necessary for the system to reach equilbrium.
Consequently, the KS regime is expected to be located in the time
region : $\tau < t < T_{relax}$. Note that, due to the condition
of Eq.(\ref{frequentmeasurement}),
$ T_{relax} \equiv 1/2 (\sigma \hbar)/(V^{2}) =1/2 (\hbar/\sigma)
(\sigma/V)^{2} >> \hbar/\sigma$.
	 
 The exact system dynamics are described by Eqs. (\ref{tutte}), which 
means 4 coupled differential equations. Using either a perturbation 
approach or a projection method~\cite{mazza}, both resting on the basic 
condition of Eq.(\ref{frequentmeasurement}), 
we reduce the number of coupled differential 
equations from 4 to 2. The solution is obtained by diagonalizing a 
two-dimension matrix, whose eigenvalues are:
	   \begin{equation}
	   \Lambda_{\pm} = \frac{1}{2} \pm \frac{1}{2}
	   exp(-\frac{2V^{2}t}{\sigma \hbar} ) .
	   \label{eigenvalues}
	   \end{equation}
	  
	   The time evolution of the von Neumann entropy
      \begin{equation}
	S(t) = - Tr \{\rho(t) \log \rho(t)\},
	\label{vonneumann}
	\end{equation}
	corresponding to the earlier approximations reads
\begin{equation}\label{exactentropy}
S(t)= -[\frac{1}{2} + \frac{1}{2} exp(-\frac{2V^{2}t}{\sigma \hbar})]
\log [\frac{1}{2} + \frac{1}{2} exp(-\frac{2V^{2}t}{\sigma \hbar})]
\end{equation}
\[
-[\frac{1}{2} - \frac{1}{2} exp(-\frac{2V^{2}t}{\sigma \hbar})]
\log [\frac{1}{2}-  \frac{1}{2} exp(-\frac{2V^{2}t}{\sigma \hbar})].
\]
Note that this time evolution has the same initial condition as that of
Eq.(\ref{harmonic}), namely, the system at the initial time is
in the state $|1\rangle$.
       
The approximated time evolution of $S(t)$ given by
Eq.(\ref{exactentropy}) corresponds to setting equal to $0$ the finite
size of the transition region. To establish the slope of the
the regime of entropy increasing linearly in time, we have to
consider a time scale of the order of $\tau$, namely $V^{2}t/(\sigma
\hbar) = (V/\sigma) << 1$ thereby
making it possible for us to replace Eq.(\ref{exactentropy})
with the following approximated expression
\begin{equation}
S(t) \approx \frac{V^{2}t}{\sigma \hbar} + \frac{V^{2}t}{\sigma
\hbar}\log (\frac{\sigma^{2}}{V^{2}}).
\label{approximatedevolution}
\end{equation}
	       Consequently, we obtain that a good candidate for the quantum
	       KS entropy is given by
	       \begin{equation}
	       \dot{S}(t) = \frac{V^{2}}{\sigma \hbar}
	       [1 +\log (\frac{\sigma^{2}}{V^{2}})].
	       \label{quantumksentropy}
	       \end{equation}
	      
	       To prove that this is really a KS entropy we decide to adopt
	       the following perspective. Following the authors of
	       Refs.\cite{carmichael,gisin,dalibard} we interpret the
	       decoherence process as a true wave-function collapse occurring
	       with the rate $\sigma/\hbar$. We can adopt the following
	       prescription. We imagine that a measurement process takes
	       place at regular intervals of time: $\tau, 2 \tau, \ldots, n
	       \tau\ldots$, where, in accordance with
the earlier definition, $\tau = \hbar/\sigma$. 
At the moment of
	       the first wave-function collapse, the system, which in the
	       absence of the measurement process would follow
	       Eq. (\ref{harmonic}), jumps to the state $|2\rangle$ with probability
	       $x \equiv |sin (V \tau /\hbar)|^{2}$ and to the state $|1\rangle$ with
	       probability $1-x =|cos (V \tau/\hbar)|^{2}$. Due to the fact that we
	       adopt the condition of Eq.(\ref{frequentmeasurement}) we set
	      \begin{equation}
		  x = (\frac{V}{\sigma})^{2}.
		  \label{probability}
		  \end{equation}
Since $x<<1$, at the first step the system will jump with a very
large probability into the state $|1\rangle$. At the next step the
probability of jumping into the same state is the same as at the
first step, and so on. Consequently, the system will collapse for
a large number of times into the state $|1\rangle$. 
However, since the probability of collapsing into the state $|2\rangle$ is small but non
vanishing, sooner or later the system will collapse into that
state, and, afterward it will keep collapsing into that state for
a large number of times. In literature this effect is called
$Zeno$ effect\cite{zeno}. If we associate the state $|1\rangle$ with $1$
and the state $|2\rangle$ with $0$, this will have the effect of
creating a symbolic sequence of $1$'s and $0$'s.
		 
Let us calculate the KS entropy of this symbolic sequence.		
According to the literature prescriptions to evaluate 
the KS entropy~\cite{beck,dorfman}, 
we have to proceed as follows:
We fix a window of size $N$, namely, containing $N$ symbols of
this sequence, and we move the window along the sequence,
recording the combinations of symbols $\omega_{0},
\omega_{1},\ldots \,\omega_{N-1}$, with $\omega_{i} = 0, 1$,
corresponding to any window position. Since the sequence is
assumed to be infinite, we can evaluate the frequency of occurence
of any given string and we can identify this frequency with the
probability $p(\omega_{0},\omega_{1},\ldots ,\omega_{N-1})$. 
The KS entropy is defined as
\begin{equation}
h_{KS} \equiv lim_{N\rightarrow \infty} \frac{H(N)}{N},
\label{ksdefinition}
\end{equation}
where $H(N)$ is the Shannon entropy:
\begin{equation}	  \label{shannon}
H(N) \equiv -\sum \limits_{\omega_{0},...,\omega_{N-1}}
p(\omega_{0},...,\omega_{N-1}) \log p(\omega_{0},...,\omega_{N-1}).
\end{equation}
		 
To make our calculation easier, let us imagine a case where 
the probability of getting $|1\rangle$, 1-$x$, and the probability of getting 
$|2\rangle$, are fixed. This produces a sequence with long strings of 1's 
interspersed with very few 0's. This is the former of the two kinds 
of sequences that we want to discuss here. The Zeno effect results in 
a different kind of sequence, the latter of the two here under 
discussion. In fact, if a collapse into $|2\rangle$ occurs, it is easily seen 
that since that moment the quantity $x$ becomes the probability for the 
wave function to collapse into $|1\rangle$ and 1 - $x$ becomes the probability 
for the wave function to collapse into $|2\rangle$. In other words, the 
latter sequence, corresponding to the realization of the Zeno effect, 
is obtained from the former by changing the 1's of the second, 
fourth, sixth string, and so on, into 0's, and by ignoring the 0's at 
the border between a given long string and the next long string.

We prove that the KS entropy of the former sequence is the 
same as the KS entropy of the latter. The evaluation of the KS 
entropy of the former sequence is done noticing that at the first 
step there is a probability $x$ of getting $0$ and a
probability $(1-x)$ of getting $1$. At the second 
step there are
four possibilities, the string $00$, with 
probability $x^{2}$,
the string $11$, with probability  $(1-x)^{2}$ , 
and the strings
$01$ and $10$, both with the same probability $x(1-x)$.
In general at the $N-th$ step there will be $2^{N}$ different
strings, with probability
		 
		  \begin{equation}
		      p^{(N)}_{k} = x^{k}(1-x)^{N-k} \quad  k = 0,\ldots , N.
		      \label{generalcase}
		      \end{equation}
		      As a consequence, we can write:
		      \begin{equation}
			  H(N) = -\sum_{0}^{N} \frac{N!}{k!(N-k)!} 
x^{k}(1-x)^{N-k}\log x^{k}(1-x)^{N-k}.
			  \label{generalexpression}
			  \end{equation}
After some algebra, Eq.(\ref{generalexpression}) is proven to be
proportional to $N$ through such a factor that the adoption of
Eq.(\ref{ksdefinition}) yields
\begin{equation}
h_{KS} = x \log (\frac{1}{x}) + (1-x) \log (\frac{1}{1-x}).
\label{lyapunovcoefficient}
\end{equation}
     It is interesting to notice that this result coincides with that
     of Dorfman, Ernst and Jacobs\cite{modifiedbernouilli} who studied
     the asymmetric Bernouilli map given by
    
      \begin{eqnarray}
   y_{n+1} = y_{n}/(1-x) \quad \mathrm{for} \quad 0 \leq y_{n} < x \nonumber \\
   y_{n+1} = (y_{n}- x)/x \quad \mathrm{for} \quad x \leq y_{n} <1 .
      \end{eqnarray}
     These authors evaluated the Lyapunov coefficient of this map and
     found it to be equal to the KS entropy of Eq.(\ref{lyapunovcoefficient}).
     In fact, we know from the
     the Pesin theorem\cite{pesin} that the KS entropy of a chaotic map
     is its Lyapounov coefficient.
    
     To prove that the theoretical prediction of
     Eq.(\ref{lyapunovcoefficient}) is really equivalent to the KS entropy 
of a Zeno effect, we create a symbolic sequence by randomly drawing a
     number of the interval $[0,1]$. 
     The drawing is done at the integer times: n=1,2,...
 If this number is smaller than the
     number $x <<1$, we select 0, if it is larger we select 1.
  The probability of selecting $1$ is $1-x$, and consequently 
it is much larger than the probability of selecting $0$. As small as 
the probability of $0$ is, if we get $0$, we reverse the procedure, 
and we associate the numbers larger than $x$ with $0$, and those 
smaller than $x$ with $1$.
 Proceeding along these lines we find the
     sequence illustrated in Fig. 1. The KS entropy of this sequence is
     then evaluated numerically using the prescription of
     Eq.(\ref{shannon}). For windows of size $N <20$ this can be
     easily done. The result is illustrated in Fig. 2, for $ x =
     0.05$. We see that the agreement between the theoretical
     prediction of Eq.(\ref{lyapunovcoefficient}) and the numerical
     result is excellent. We can thus conclude that the
     asymmetric map of Dorfman, Ernst and
     Jacobs\cite{modifiedbernouilli} is a dynamic system
     statistically equivalent to the Zeno effect\cite{zeno}.
    
Note that to make complete the connection with the quantum measurement
process, we have to move from the unit time picture adopted to
derive Eq.(\ref{lyapunovcoefficient}) to
a picture expressed in terms of physical time. To do that we have 
to notice that
in the physical process under study the elementary
time step is, as we have seen, $\tau = \hbar/\sigma$. We
replace $x = (V/\sigma)^{2}$ into
Eq.(\ref{lyapunovcoefficient}), make the approximation $\log(1-x)
\approx -x$ and multiply the result by $\sigma/\hbar$. The
result is:
\begin{equation}
h_{KS} = \frac{V^{2}}{\sigma \hbar} [1 +
\log(\frac{\sigma^{2}}{V^{2}})],
\label{centralresult}
\end{equation}
which coincides with Eq.(\ref{quantumksentropy}), which is 
the rate of the physical
entropy in the time
regime that the earlier theoretical arguments have identified with
the quantum KS regime.

In addition to the benefit, of merely conceptual interest, of
extending the correspondence between the physical entropy and the KS
entropy also to cases with no classical counterpart, the results of
this letter, as simple as they are, makes it possible to derive the
results about fluorescence spectra in a single-atom double-resonance
experiment\cite{cook} using methods borrowed from the literature on
intermittent processes\cite{geisel}. In fact, when the condition
$x << 1$ applies, the left hand part of the asymmetric map of Eq.(16)
becomes equivalent to the continuous time equation
\begin{equation}
\dot{y} = [x/(1-x)] y(t)
\label{intermittency}
\end{equation}
ranging from $y = 0$ to $y=1$. 
Note that Eq.(\ref{intermittency}) applies to the whole interval [0,1] 
but the point y = 1. 
Thus the point y = 1 and the remainder part of the interval 
[0,1] play the same role as the chaotic region and the laminar region 
of the intermittent map of Ref.~\cite{geisel}, respectively. 
As done in Ref.~\cite{geisel}, 
we assume that the injection of the trajectory back into the 
laminar region is uniform. Thus, using Eq.(\ref{intermittency}) 
and adopting the same 
approach as that of Ref.~\cite{geisel}, 
we evaluate the distribution of 
sojourn times, $\psi(t)$. This is:

\begin{equation}
\psi(t) = exp(-[x/(1-x)]t).
\label{waitingtimedistribution}
\end{equation}
On the other hand, we know\cite{allegro} that 
the second-order time
derivative of the correlation
function of the telegraphic noise $C(t)$ is 
proportional to the
waiting time distribution $\psi(t)$. From
Eq.(\ref{waitingtimedistribution}) we 
immediately derive,in a full agreement with the theory of
Ref.\cite{cook}, that also
this function is exponential\cite{note}.

%The adoption of the intermittency point of 
%view implies that
%the observation time $T$ be referred to the 
%mean sojourn in the
%laminar time $\langle t \rangle$ .
%It must be stressed, as an another possible 
%benefit of the very
%simple calculations illustrated in this 
%letter might be a new
%perspective of the cases where $\psi(t)$ is 
%an inverse power law:

This paper rests on a perspective which might depart from that of the 
KS entropy. Here we are imagining that a genuine random process is 
acting at regular intervals of time, $n$. The "time" $n$ is a sort of 
internal time, and the physical entropy is expected to increase 
linearly with it. The observation time T is easily related to the 
internal time through $n=\mathrm{T/\langle t \rangle}$, 
where $\mathrm{\langle t \rangle}$ denotes the mean time of 
sojourn in the laminar region. This perspective might lead to a 
departure from the KS entropy in the cases where $\psi(t)$ is the 
inverse power law:
\begin{equation}
\lim_{t \rightarrow \infty} \psi(t) = 
\frac{const}{t^{\mu}},
\;\mathrm{with} \; \mu > 1.
\label{asymptotic}
\end{equation}
We know that this inverse power law 
asymptotic behavior can be
obtained with a slow modulation of the 
rate $x$\cite{rocco}. In
the case where $2 < \mu <3$ one would 
expect that an ordinary
KS entropy emerges\cite{gaspard}. 
However, the adoption of the
KS perspective seems to imply an average 
of the fluctuating
probability $x/(1-x)$.
This paper suggests that the rate of increase of the physical entropy 
is proportional to T/$\langle t \rangle$. 
This implies that we have to make the 
average of $(1-x)/x$
rather than that of $x/(1-x)$. 
We think that all this might raise 
interesting new questions on the meaning of the KS entropy.

\newpage

\begin{figure}[r]
\centerline{
\vbox{
\psfig{figure=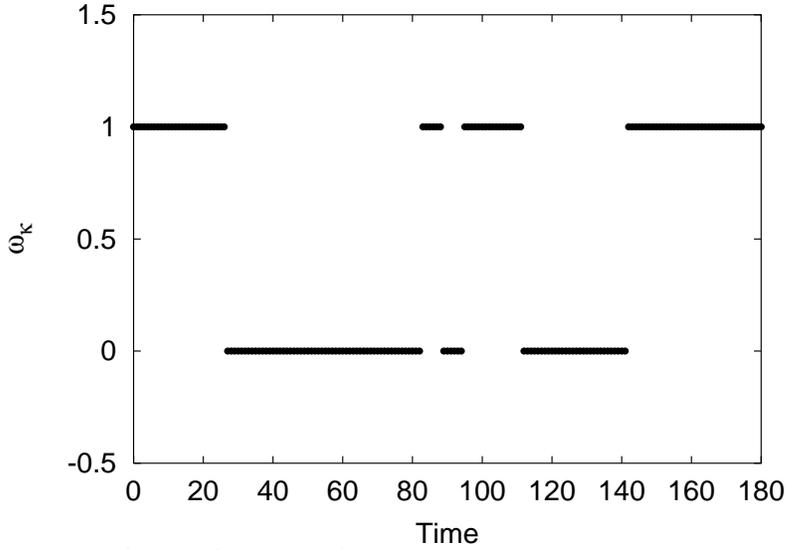,width= 4.3 in}
}}
\caption[]{
A sample of the sequence generated using the method of random 
drawing of the numbers of the interval $[0,1]$. The sequence refers 
to the case $x= 0.05$.
Time is expressed in terms of the "integer time" $n$.
}
\label{uno}
\end{figure}

\begin{figure}[r]
\centerline{
\vbox{
\psfig{figure=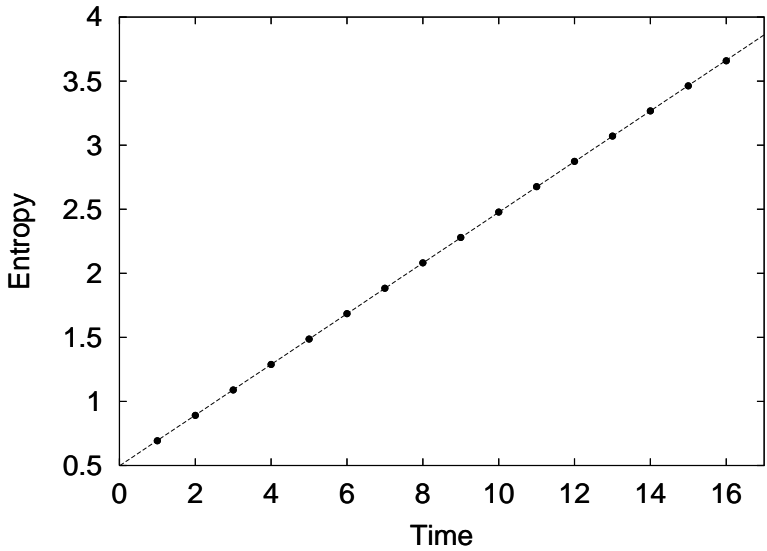,width= 4.3 in}
}}
\caption[]{
The entropy $H(N)$  as a function of $N$. The dots denote the 
results of the numerical analysis made using a moving window of size 
$N$ to evaluate the frequency p($\omega_{0},\omega_{1},...,\omega_{N-1}$). 
This probability is then adopted  to define $H(N)$ using the 
prescription of Eq.(\ref{shannon}). The dashed line denotes a straight line 
whose slope is given by the theoretical prediction of Eq.(\ref{lyapunovcoefficient}). 
The abscissa denotes the window size $N$.}
\label{due}
\end{figure}


\begin{references}
\bibitem{note} Note that the authors of Ref.\cite{cook} considered a
case where the waiting time distribution in one state can be different
from that of the other state. Thus, we have to compare our results to
the theoretical predictions of Ref.\cite{cook} in the symmetric case.
\end{references}
\end{document}